\documentclass[twocolumn,secnumarabic,amssymb,nobibnotes,aps,prx]{revtex4-1}
\usepackage{graphicx}  
\usepackage{dcolumn}   
\usepackage{bm}        
\usepackage{amssymb}   
\usepackage{epstopdf}

\newcommand{\units}[1]{\mbox{ }\mbox{#1}}

\newcommand{\ind}{\hspace{0.2 in}}

\newcommand{\textss}[1]{\scriptsize \mbox{#1}}

\newcommand{\vecbf}[1]{\mathbf{#1}}

\newcommand{\maxsamplecurrent}{5 nA }
\newcommand{\frep}{88~MHz }
\newcommand{\minsideamp}{$6 \times 10^{-4}$}

\begin{document}

\title{Ultrafast extreme ultraviolet photoemission without space charge}%

\author{Christopher Corder$^{1\dagger}$, Peng Zhao$^{1\dagger}$, Jin Bakalis$^{1}$, Xinlong Li$^{1}$, Matthew D. Kershis$^{2}$, Amanda R. Muraca$^{1}$, Michael G. White$^{1,2}$, and Thomas K. Allison$^{1}$}
\email[T. K. Allison:]{thomas.allison@stonybrook.edu \\ $^{\dagger}$ C. Corder and P. Zhao contributed equally to this work.}
\affiliation{$^{1}$Stony Brook University, Stony Brook, NY 11794-3400}
\affiliation{$^{2}$Brookhaven National Laboratory, Upton, New York 11973}
\date{\today}%
\begin{abstract}

Time- and Angle-resolved photoelectron spectroscopy from surfaces can be used to record the dynamics of electrons and holes in condensed matter on ultrafast time scales. However, ultrafast photoemission experiments using extreme-ultraviolet (XUV) light have previously been limited by either space-charge effects, low photon flux, or limited tuning range. In this article, we describe space-charge-free XUV photoelectron spectroscopy experiments with up to \maxsamplecurrent  of average sample current using a tunable cavity-enhanced high-harmonic source operating at \frep repetition rate. The source delivers $ > 10^{11}$ photons/s in isolated harmonics to the sample over a broad photon energy range from 18 to 37 eV with a spot size of $58 \times 100 \; \mu$m$^2$. From photoelectron spectroscopy data, we place conservative upper limits on the XUV pulse duration and photon energy bandwidth of 93 fs and 65 meV, respectively. The high photocurrent, lack of space charge distortions of the photoelectron spectra, and excellent isolation of individual harmonic orders allow us to observe the laser-assisted photoelectric effect with sideband amplitudes as low as \minsideamp, enabling time-resolved XUV photoemission experiments in a qualitatively new regime.

\end{abstract}
\maketitle

\section{Introduction} 
 
Angle-resolved photoelectron spectroscopy (ARPES) using synchrotron radiation has become an essential tool for condensed matter physics and surface science. The high spectral brightness of synchrotron radiation allows photoelectron spectra to be recorded with photocurrents in the nano-Ampere range. These large photocurrents, parsed by sophisticated electron energy analyzers \cite{Medjanik_NatMat2017, Jozwiak_SRN2012, Iwasawa_JSR2017}, enable detailed studies of the electronic structure of solids and surfaces in energy, momentum, and spin. Tuning the photon energy throughout the extreme ultraviolet (10-100 eV) allows experiments to map the energy dispersion relation for momentum perpendicular to the surface ($k_z$), interpret the contributions of final state effects to the measured energy distribution curves (EDC), and choose between increased surface or bulk sensitivity~\cite{Hufner_book2003}.

\ind Shortly after the development of high-power femtosecond lasers and discovery of high order harmonic generation (HHG), in which a broad range of laser-harmonics are coherently emitted from a field-ionized medium \cite{Brabec_RMP2000}, HHG was applied to surface photoemission experiments \cite{Haight_RSI1994, Bauer_PRL2001, Haarlammert_CurrOpSSMS2009}. Indeed, the range of photon energies typically emitted from HHG driven by Ti:Sapphire lasers in noble gasses is nicely coincident with the range of photon energies used by ARPES beamlines at synchrotrons. In addition to the dramatically reduced cost and footprint compared to a synchrotron source, the HHG pulses had the advantage that they could be orders of magnitude shorter than the $\sim$100 ps pulse durations of synchrotrons, enabling ultrafast time-domain studies. However, it also became immediately apparent that photoemission experiments using HHG would be drastically limited compared to what is possible at synchrotrons \cite{Haarlammert_CurrOpSSMS2009, Mathias_Collection2010}.

\ind The principle limitation comes from the so-called ``vacuum space-charge" effect \cite{Hellmann_PRB2009}. Consider a synchrotron experiment illuminating the sample with $\sim 10^{12}$ photons/second causing $\sim 10^{11}$ electrons/second (16 nA) to be emitted from the surface. The synchrotron photon pulses arrive at $\sim 100$ MHz repetition rate, so each burst of electrons emitted concurrently from a single pulse contains only $\sim 1000$ electrons. In contrast, due to the high peak powers required to drive the HHG process efficiently, HHG is typically restricted to $<$100 kHz repetition rates. In order to maintain the same photocurrent, the electrons must then be concentrated by more than 1000 times, to more than $10^6$ electrons/pulse. The charging of space  at these electron densities distorts the photoelectron spectrum on the eV energy scale, whereas synchrotron beamlines now routinely record photoelectron spectra with meV resolution \cite{Reininger_AIP2007}. Practitioners of time-resolved photoelectron spectroscopy using HHG are then forced to compromise on the applied photon flux, focused spot size, resolution, fidelity of the signal, or some combination thereof \cite{Plotzing_RSI2016, Mathias_Collection2010, Al-Obaidi_Thesis2016}. Compounding the problem is the fact that a time-resolved pump-probe experiment inherently requires much more data than a static one, since one must record spectra at many pump-probe delays and only a small fraction of the sample's electrons should be excited by the pump. 

\ind With the constraint of space charge setting the fundamental limits on the performance of photoemission experiments, this phenomenon has been extensively studied over a wide range of electron kinetic energies and pulse durations, both experimentally and theoretically \cite{Hellmann_PRB2009, Plotzing_RSI2016, Frietsch_RSI2013, Mathias_Collection2010, Graf_JApplPhys2010, Passlack_JApplPhys2006, Zhou_JElecSpec2005}. For sub-ps pulses and electron kinetic energies in the $\sim$ 5-100 eV range produced from conductive samples, both shifts and broadening of the photoelectron spectra features are observed to scale with linear electron density $\rho \equiv N/D$, where $N$ is the number of electrons emitted from the sample per pulse and $D$ is the spot size of the light on the sample. Expressed in terms of the average sample current ($I_{\textss{sample}}$) and repetition rate ($f_{\textss{rep}}$), this gives:

\begin{equation}\label{eqn:sc}
\Delta E_{\textss{s,b}} = m_{\textss{s,b}} \frac{I_{\textss{sample}}}{e f_{\textss{rep}} D}
\end{equation}

where $\Delta E_{\textss{s,b}}$ is the energy shift (s) or energy broadening (b) of the photoelectron spectrum, $e$ is the charge of the electron, and $m_{\textss{s,b}}$ are empirical scaling factors. Reports of the slope parameters $m_{\textss{s,b}}$ in the literature have varied by a factor of 2 \cite{Hellmann_PRB2009, Plotzing_RSI2016, Frietsch_RSI2013, Mathias_Collection2010}. Recently, Pl\"otzing et al. \cite{Plotzing_RSI2016} have studied these effects for multiple spot sizes and determined $m_b = 2.1 \times 10^{-6}$ eV-mm and and $m_s = 3.2 \times 10^{-6}$ eV-mm with an estimated systematic uncertainty of less than 20\%. 

\begin{figure}[t!]
\begin{center}
\includegraphics[width = 9.2 cm]{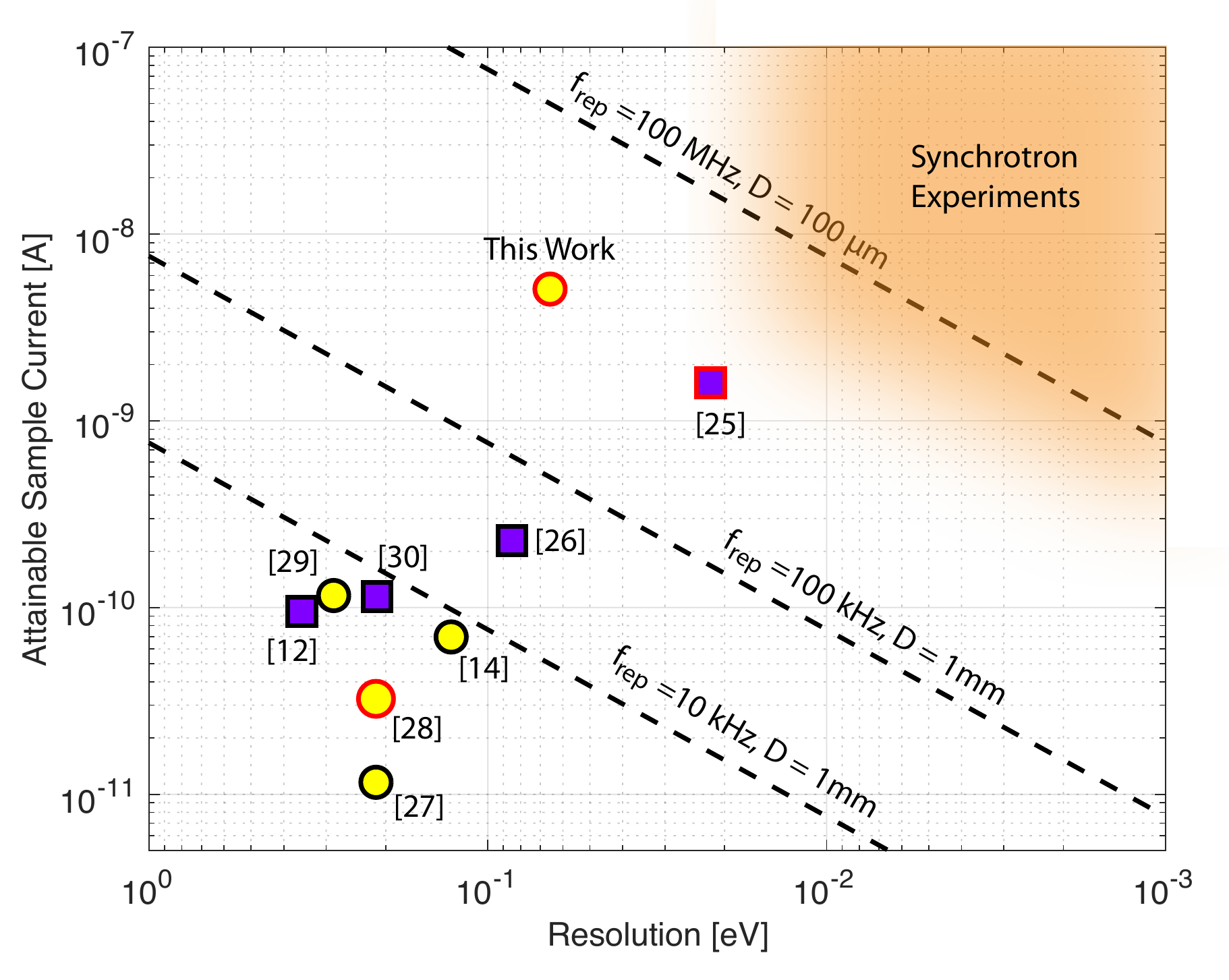} 
\caption{Dashed lines are created by evaluating Eq.~\ref{eqn:sc} for different repetition rates assuming 1 mm spot size and symbols represent published results applying HHG to surface photoemission. Yellow circles represent results from tunable HHG systems and purple squares represent setups where the photon energy is not tunable {in-situ}. Symbols with black edges represent space-charge limited spectrometers, and symbols with red edges represent systems that are not yet space-charge limited. For space-charge-limited systems, the $(x,y)$ positions represent the case where space charge broadening and the photon bandwidth add equally in quadrature. See appendix~\ref{ap:fig1} for detailed explanation on how symbol placement was calculated based on published results \cite{Mills_CLEO2017,*Jones_Private,Buss_SPIE2017,Frietsch_RSI2013,Artemis_Private,Chiang_NJP2015,Plotzing_RSI2016,Ojeda_SD2016,EICH_JElecSpec2014}.}
\label{fig:sourcecomparison}
\end{center}
\end{figure}

\ind Figure \ref{fig:sourcecomparison} illustrates the constraints on attainable sample current for a given resolution according to Eq.~(\ref{eqn:sc}). The dashed lines indicate the space charge limits for sub-ps laser-based systems of different repetition rates assuming a 1 mm spot size - large by ARPES standards. Even at the high repetition rate of 100 kHz and the coarse resolution of 100 meV, space charge constraints still limit the sample current to 760 pA. These low data rates then often restrict experiments to strongly excited samples using absorbed fluences on the order $\sim$ 1 mJ/cm$^2$ \cite{Mathias_NatComm2016, Borgwardt_JPhysChemC2015, Saathoff_PRA2008}. At these fluences ultrashort pump pulses also produce many electrons through multiphoton processes which add to the space-charge problem \cite{Ultstrup_JElecSpec2015, Oloff_JApplPhys2016, Borgwardt_Thesis2016}. For non-conducting samples such as insulators or liquids, where positive charges left behind in the sample are immobile, shifts of the photoelectron spectra can be much larger and show a non-trivial delay dependence in pump-probe experiments which can be difficult to separate from the dynamics of interest \cite{Al-Obaidi_NJP2015}.  

\ind Motivated by the inverse dependence of Eq.~(\ref{eqn:sc}) on $f_{\textss{rep}}$, in this article we demonstrate the application of a widely tunable cavity-enhanced high-harmonic generation (CE-HHG) light source \cite{Mills_JPhysB2012} to the difficult problem of time-resolved surface photoemission. By performing experiments with high flux at \frep repetition rate, nanoamperes of sample current can be generated from a sub-100 micron laser spot with space charge effects less than 10 meV, comparable to synchrotron-based ARPES experiments \cite{Zhou_JElecSpec2005, Hoesch_RSI2017}. As we show, this enables time-resolved photoelectron spectroscopy in a qualitatively different regime of resolution and pump-fluence than space-charge limited systems. In section \ref{sec:lightsource}, we describe critical and unique details of the light source along with its performance. In section \ref{sec:photspec} we demonstrate both static and time-resolved photoelectron spectroscopy measurements with the high dynamic range enabled by nA space-charge free sample currents. In section \ref{sec:conclusions} we discuss the comparison of this work to previous efforts and describe how the system can be further improved.

\section{Light Source and Beamline}\label{sec:lightsource}

\begin{figure*}[t!]
\begin{center}
\includegraphics[width = 0.95\textwidth]{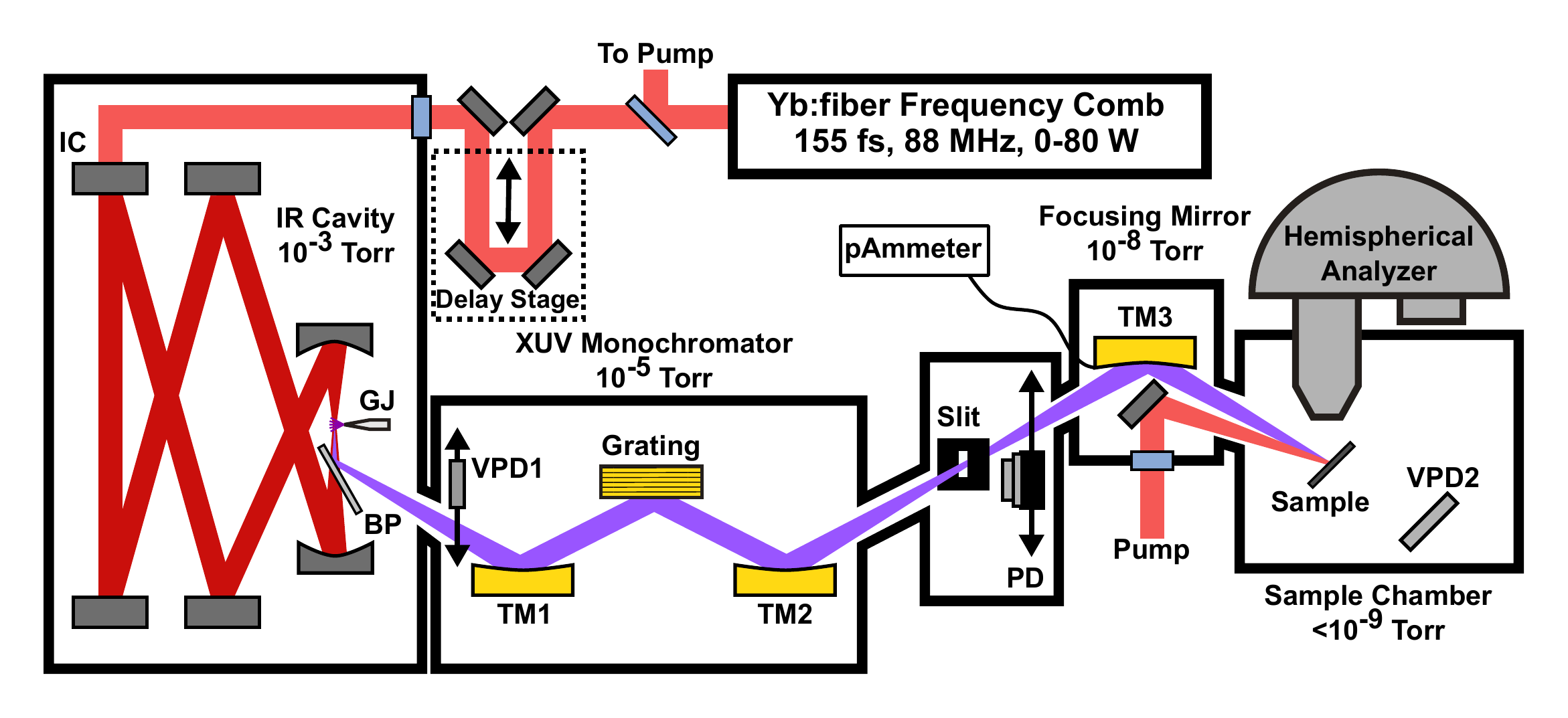} 
\caption{CE-HHG source and beamline. High-order harmonics of a resonantly enhanced Yb:fiber frequency comb are generated at the focus of a 6 mirror enhancement cavity and coupled into an XUV beamline. A pulse-preserving monochromator selects one harmonic which is focused on a sample under UHV conditions. BP = Brewster plate, VPD = vacuum photodiode, TM = toroidal mirror, PD = XUV photodiode, GJ = gas jet, IC = input coupler.}  
 
\label{fig:setup}
\end{center}
\end{figure*}

The experimental setup is shown in Fig.~\ref{fig:setup}. A home-built 80 W, 155 fs frequency comb laser with a repetition rate of \frep and and a center wavelength of 1.035~$\mu$m ($h \nu = 1.2$ eV) is passively amplified in a 6 mirror enhancement cavity with a 1\% transmission input coupler. We have described the laser in detail previously \cite{Li_RSI2016}. The laser is locked to the cavity using a two-point Pound-Drever-Hall lock as described in \cite{Foltynowicz_APB2013, Corder_SPIE2018}. Harmonics are generated at a 24 $\mu$m FWHM intracavity focus and reflected from a sapphire wafer placed at Brewster's angle for the resonant 1.035 $\mu$m light. Noble gasses are injected to the focus using a fused silica capillary with a 100 $\mu$m inside diameter. We optimize the nozzle position by moving it to maximize the photocurrent observed on a stainless steel vacuum photodiode (VPD1) \cite{Feuerbacher_JApplPhys1972}. Typical photocurrents from VPD1 are in the range of 100 to 300 nA. When generating harmonics, we also dose each intracavity optic with a mix of ozone and O$_2$ from a commercial ozone generator to prevent hydrocarbon contamination allowing continuous operation. 

\ind Typical intracavity powers for generating harmonics range from 5-11 kW, depending on the generating gas and desired harmonic spectrum, corresponding to intracavity peak intensities in the range of 0.6 to 1.3 $\times 10^{14}$ W/cm$^2$. Intracavity nonlinear effects are observed from both the HHG gas and self-phase modulation in the Brewster plate, dropping the cavity's power enhancement and necessitating careful tuning of servo loop offsets \cite{Allison_PRL2011, Yost_OptExp2011, Corder_SPIE2018}. For example the power enhancement drops from 270 at low power to 200 
at 7.5 kW of intracavity power when generating harmonics in krypton. 

\ind The outcoupled harmonics are collimated by a 350 mm focal length toroidal mirror at 3 degrees grazing angle (TM1) that forms the first part of a single off-plane grating pulse-preserving monochromator similar to the design of Frassetto et al. \cite{Frassetto_OptExp2011}. The harmonics strike a motorized grating at a 4 degree grazing angle and are refocused by a second $f=350$ mm toroidal mirror (TM2) at an adjustable slit. For all data presented here, the monochromator grating has 150 grooves/mm and is blazed for optimum diffraction efficiency for $\lambda = 35$ nm. With this grating the monochromator selects an individual harmonic with tolerable pulse broadening but does not narrow the transmitted harmonic bandwidth. The exit slit plane of the monochromator is 1:1 imaged to the sample using another 350 mm focal length toroid at 3 degrees grazing angle (TM3). Mirror TM3 is electrically floated such that the photocurrent of electrons ejected from the mirror surface can be used as a passive XUV intensity monitor. All beamline optics are gold coated and the XUV light is polarized perpendicular to the plane of incidence (s-polarization).

\begin{figure}[t!]
\begin{center}
\includegraphics[width = 8.5 cm]{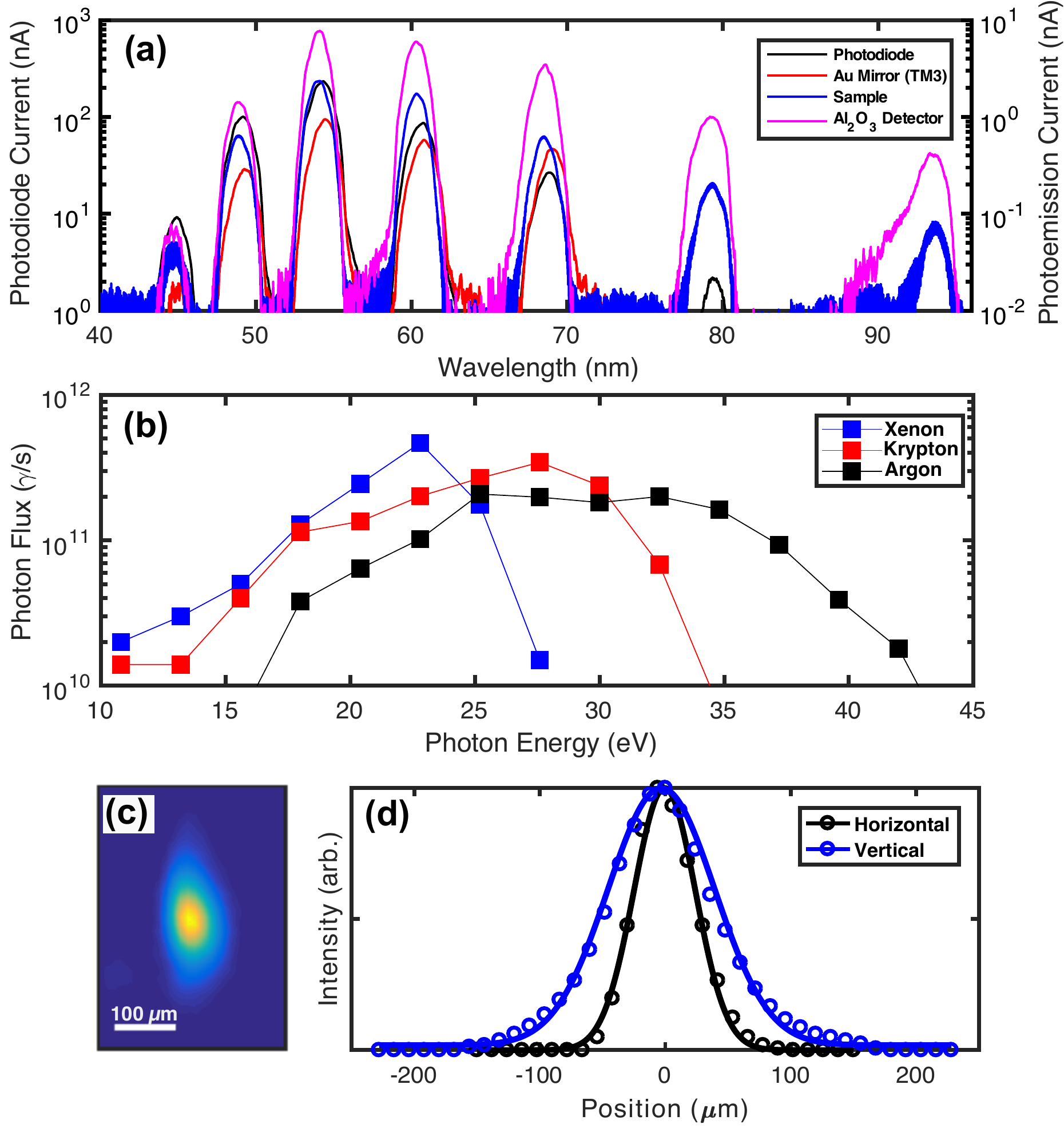} 
\caption{a) An HHG spectrum from xenon gas measured by rotating the monochromator grating while using the four detectors after the exit slit. The photodiode current (black) uses the left y-axis whereas the the photoemission current from three downstream surfaces uses the right y-axis. Note that the harmonic linewidths are not resolved with the pulse-preserving monochromator design. b) The photon flux delivered to the sample for each harmonic generated with the three gases. The flux has been calibrated using literature values for quantum efficiencies and no corrections for mirror losses have been made. c) The 27th harmonic from Ar imaged with a Ce:YAG crystal at the sample position. d) Lineouts through the centroid of (c) fit with Gaussian functions demonstrating 58 $\mu$m x 100 $\mu$m spot size (FWHM).}  
 
\label{fig:HHGflux}
\end{center}
\end{figure}

\ind We detect the XUV flux exiting the monochromator and delivered to the sample using four separate detectors: an aluminum coated silicon photodiode (PD, Optodiode AXUV100Al), the photocurrent from TM3, the photocurrent from the sample, and the photocurrent from an Al$_2$O$_3$ vacuum photodiode \cite{CANFIELD_NBST1987} (VPD2) placed at the end of the surface science chamber. Figure \ref{fig:HHGflux}a) shows a typical HHG spectrum from xenon gas measured using each of the four detectors as the monochromator grating is rotated. The observed harmonic linewidths in Fig.~\ref{fig:HHGflux}b) are due to the intentionally small resolving power of the pulse-preserving monochromator, not the intrinsic harmonic linewidth. 

\ind The photon flux can be calculated using the measured photocurrent from all the detectors and literature values for the quantum efficiencies. All of these separate calculations agree within a factor of 2. Since contamination and surface oxidation only cause the quantum efficiency of XUV detectors to decrease, all calculated photon fluxes represent lower limits. In Fig.~\ref{fig:HHGflux}b), the higher of the two lower limits from the PD or VPD2 are plotted as a function of photon energy and for three different generating gasses: argon, krypton, and xenon. As can be seen in Fig.~\ref{fig:HHGflux}b), even using a single monochromator grating, by changing the generating gas, a flux of more than $10^{11}$ photons per second is delivered to the sample over a broad tuning range. At lower photon energies, the higher efficiency of generation in Kr and Xe compensates the reduced diffraction of efficiency of the grating blazed for 35 eV. Higher fluxes can be obtained at lower photon energies using different gratings. For example, we have observed $7\times 10^{11}$ photons/s in the 21st harmonic from krypton ($h \nu = 25.1$ eV) using a 100 groove/mm grating blazed for 55 nm. These fluxes are within one order of magnitude of what is available from many state-of-the-art synchrotron beamlines dedicated to ARPES \cite{Hoesch_RSI2017, Reininger_AIP2007,  Iwasawa_JSR2017}. Critically, since at 88 MHz, $7 \times 10^{11}$ photons/s corresponds to only 8,000 photons/pulse - also comparable to synchrotrons - all of this flux is usable for high-resolution photoemission experiments.  

\ind To measure the XUV spot at the sample, we image the fluorescence from a Ce:YAG scintillator plate placed at the sample position. Figure \ref{fig:HHGflux}c) shows the image and Fig.~\ref{fig:HHGflux}d) shows Gaussian fits in the both the horizontal and vertical along lineouts corresponding to the image centroid. The data indicate a clean elliptical beam with a FWHM of 58 $\mu$m in the horizontal 100 $\mu$m in the vertical. Also, we measure that approximately 70\% of the XUV light can be transmitted through a 100 $\mu$m diameter pinhole oriented at 45 degrees to the beam axis. This spot size is again similar to what is used at synchrotron beamlines \cite{Hoesch_RSI2017, Reininger_AIP2007}. When comparing to previous HHG results it is important to note that in our case this small spot size and high flux are actually usable for experiments due to the absence of space-charge effects at \frep repetition rate. A small spot size enables studying spatially inhomogeneous samples (for example produced by exfoliation \cite{Huang_ACSNano2015}), requires less pump-pulse energy in pump/probe experiments, and is necessary for achieving high angular resolution in ARPES.

\begin{figure}[t!]
\begin{center}
\includegraphics[width = 8.5 cm]{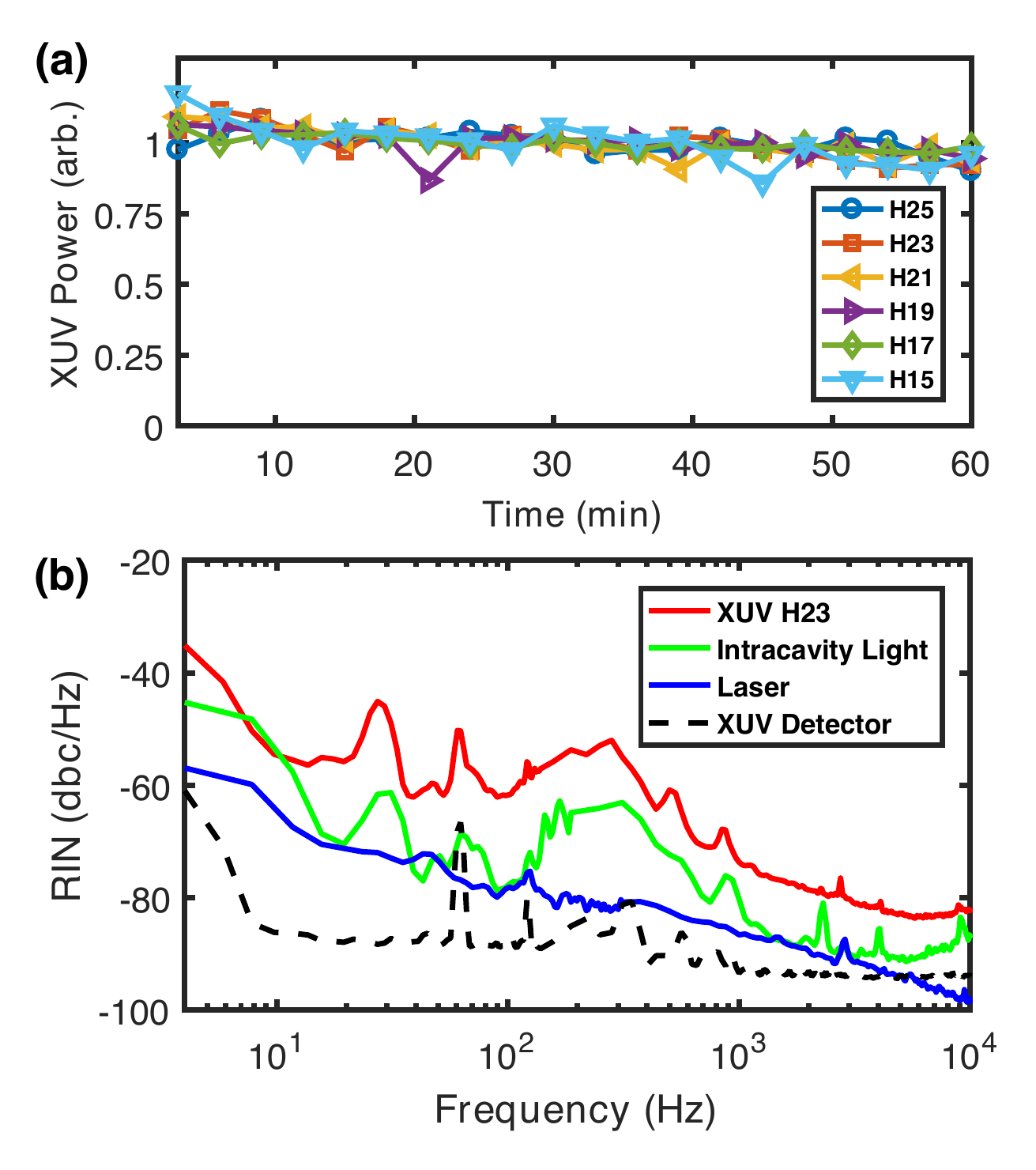} 
\caption{a) Normalized XUV flux measured with PD after the monochromator for harmonics 15-25 from Kr over 1 hour without human intervention. c) Relative intensity noise (RIN) of the 23rd harmonic (red), intracavity laser light (green), and Yb:fiber laser (blue), along with the detector noise floor (black dashed).}
 
\label{fig:spotstable}
\end{center}
\end{figure}

\ind We evaluated the amplitude noise and long term stability of the system using the PD detector. For the long term stability, we recorded a series of monochromator scans over a one hour period without any human tuning of the laser alignment or servo loop. Figure \ref{fig:spotstable}a) shows the relative power in the harmonics from Kr with a delivered flux greater than $10^{11}$ photons/s measured every 3 minutes. The RMS fluctuations averaged over all the harmonics over this period are 5\%. Similar results are obtained for HHG in Ar or Xe as well. On longer time-scales, slow drifts in the laser alignment into the cavity and servo-loop offsets require occasional tuning to maintain the flux levels at those shown in Fig.~\ref{fig:HHGflux}b). It is also important to note that since more than 100 pA of photocurrent is observed from TM3, drifts in the XUV flux can also be normalized using this in-situ monitor, as is commonly done at synchrotrons. At the time of writing we have run the source on a near-daily basis without venting the vacuum system or performing any alignment of the in-vacuum optics for more than 2 months with stable and reproducible results, enabling the photoelectron spectroscopy experiments discussed in the next section. 

\ind For pump-probe experiments it is often advantageous to use lock-in detection to extract small signals from large backgrounds. Fig.~\ref{fig:spotstable}b) shows the amplitude noise (RIN) of the 23rd harmonic from Kr measured using the photodiode current amplified with a transimpedance amplifier and recorded with an FFT spectrum analyzer. For frequencies above 400 Hz, the RIN level is below -60 dBc/Hz, which can enable small differences in the photoelectron spectra to be recorded via lock-in detection. At this noise level, EDC or ARPES signals up to $10^6$ counts/second/bin can be photoelectron-shot-noise limited with proper correction for drift using the TM3 photocurrent.

   
\section{Photoemission}\label{sec:photspec}

\ind Photoelectron spectroscopy measurements are preformed under ultra-high vacuum conditions in a surface science endstation equipped with a hemispherical electron energy analyzer (VSW HA100). The analyzer is specified to have an angular acceptance of $\pm$ 4 degrees at the input and has a channeltron detector at the exit. 
The endstation is also equipped with a sputter gun, a LEED, a quadrupole mass spectrometer, an Al K$\alpha$ x-ray source, and a sample manipulator that can be cooled and heated between 100 and 1000 K. Also mounted on the sample manipulator are the Ce:YAG scintillator and pinhole mentioned in section \ref{sec:lightsource}. For all data presented here, the sample is oriented normal to the analyzer axis and 45 degrees to the XUV beam. The electric field vector of the XUV light is in the plane of incidence (p-polarized) and the analyzer axis. 

\begin{figure}[t!]

\includegraphics[width = 8.0 cm]{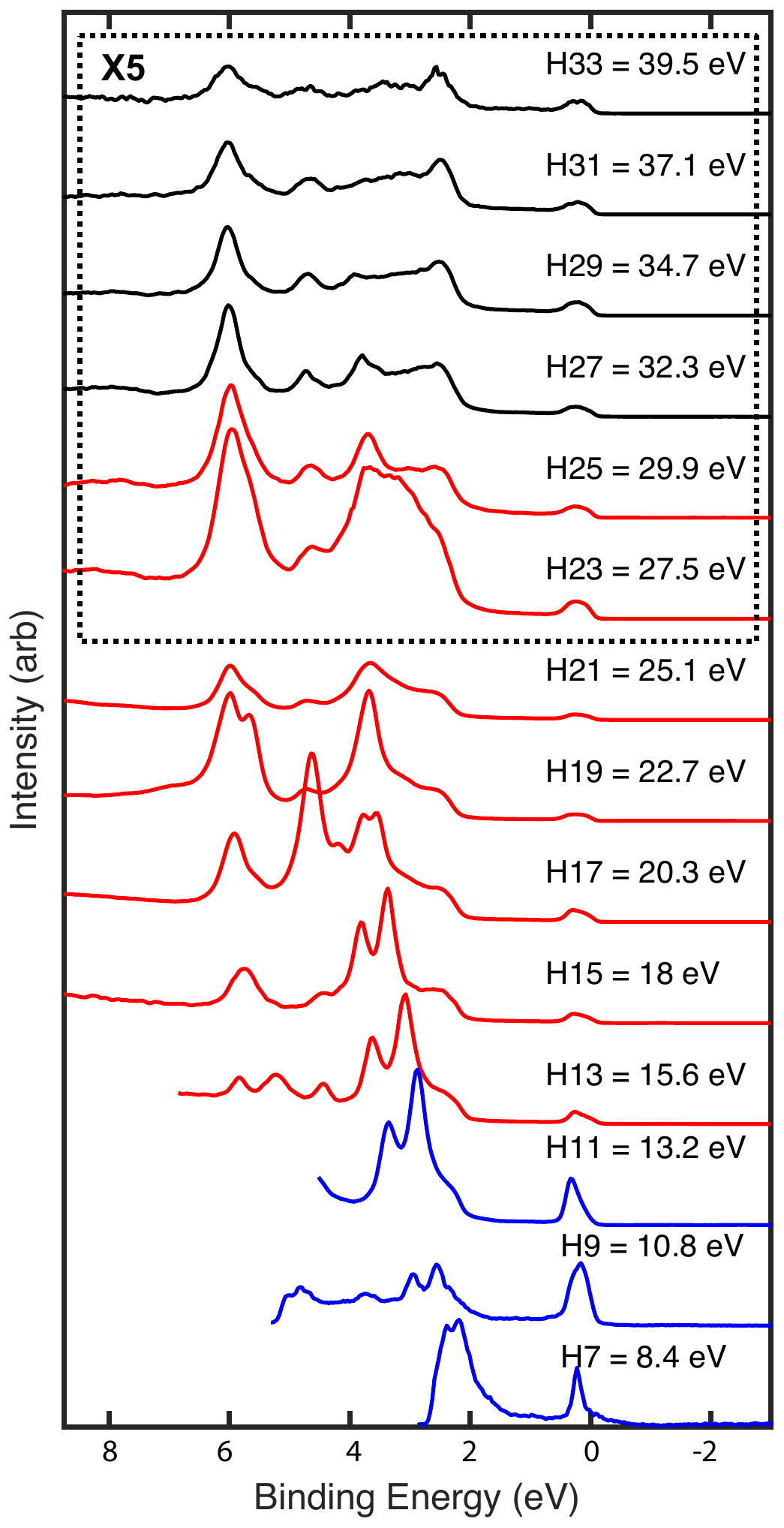} 
\caption{Static photoelectron spectra of a Au (111) surface taken with harmonics 7 through 33, vertically offset for clarity. The color indicates the gas used to generate the harmonic; Ar (black), Kr (red), Xe (blue). Each EDC is normalized to the photocurrent measured at TM3, and spectra taken with photon energies above 25.1 eV have been enlarged by $\times5$.}  
 
\label{fig:allUPS}

\end{figure}

\ind Figure \ref{fig:allUPS} shows photoelectron spectra from an Au~(111) surface at 100 K temperature obtained using each harmonic between the 7th ($h \nu = 8.4$ eV) and 33rd ($ h \nu $ = 39.5 eV). Each spectrum was acquired with 34 meV steps individually measured with 1 second of integration for a total scan time of $\sim$6 minutes or less. 
At the d-band peaks, the electron count rates can exceed 1 MHz. These static spectra are in good agreement with those recorded by Kevan et al. \cite{Mills_PRB1980,Kevan_PRB1987} using tunable synchrotron radiation. The clearly visible dispersion of the d-bands at binding energies between 3 and 7 eV and the large photon energy dependence of the relative amplitudes of the peaks highlight the importance of conducting photoemission experiments with a tunable source. The same final state effects also strongly influence time-resolved photoelectron spectra and tunability should be considered no less important, as has been emphasized by previous authors \cite{Zhu_JElecSpec2015}.

\ind The resolution of the setup can be determined by analyzing the sharpness of the Fermi edge and is dominated by the energy analyzer. Fermi edge widths as low as 110~meV are measured depending on alignment. The best resolution we have been able to observe in any photoemission experiment using this analyzer is 89 meV using a He I lamp and a Kr gas target. From this data we can place a conservative upper limit on the single harmonic photon energy bandwidth of $\sqrt{(110 \units{meV})^ 2 - (89 \units{meV})^2}$ = 65 meV. We have several reasons to believe that the single harmonic linewidth is substantially lower than this. First, the observed Fermi edge sharpness is  found to be completely independent of the HHG generating conditions or harmonic used under variation of a large range of parameters. For example, reducing the driving laser intensity or using the cavity to narrow the bandwidth of the driving pulse \cite{Foltynowicz_APB2013} should both reduce the harmonic bandwidth, but no change in the the photoelectron spectrum is observed. Furthermore, Mills et al. \cite{Mills_CLEO2017,*Jones_Private} have reported single harmonic linewidths from a cavity-enhanced HHG source similar to ours, but using even shorter driving pulses, with FWHM as low as 32 meV. Starting with longer driving pulses in our setup, we expect harmonic linewidths narrower than this are obtainable. Further investigation of the energy resolution will be the subject 
of future work as we upgrade our electron energy analyzer.

\begin{figure}[t!]
\begin{center}
\includegraphics[width = 8.5 cm]{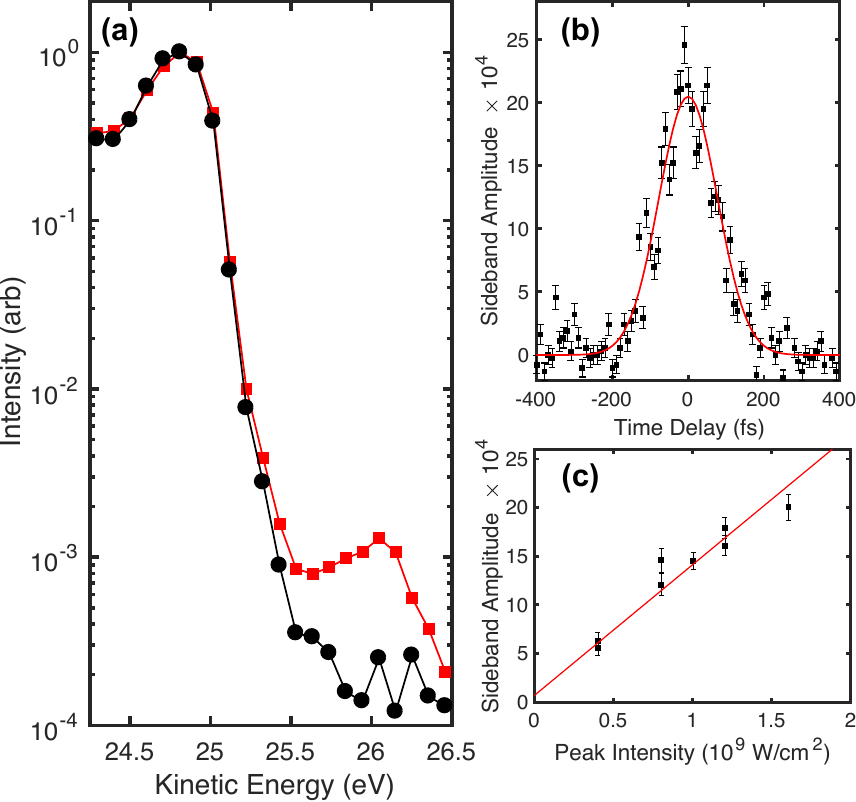}
\caption{a) The photoelectron spectrum of the Au (111) Fermi edge taken without (black) and with a 1.035 $\mu$m pump pulse (red) at a peak intensity of 1.3$\times 10^9$ W/cm$^2$. A LAPE sideband of the surface state peak at 24.8 eV is observed at 26 eV. b) The magnitude of the sideband at a kinetic energy of 26 eV as a function of pump probe time delay. The cross-correlation has a FWHM of 181 fs. c) The amplitude of the sideband at 26 eV as a function of pump peak intensity. A fit to the data gives a slope of $1.34\times10^{-12}$ cm$^2$/W.}  
 
\label{fig:TRUPS}
\end{center}
\end{figure}

\ind Even with the current analyzer-limited resolution, we demonstrate here that the absence of space-charge allows for time-resolved photoemission experiments that are both qualitatively and quantitatively different than what is done with space-charge limited systems. Figure \ref{fig:TRUPS} shows two photoelectron spectra near the Fermi edge of the Au (111) on a logarithmic scale, one with and another without a parallel polarized 1.035 $\mu$m wavelength laser excitation overlapped in space and time. The spectra were taken with 3 nA of sample current, or approximately 215 electrons/pulse. Consider first the black curve taken with the pump laser off. For a 100 kHz system with our spot size (or even somewhat larger), this sample current would result in broadening and shifting of the Fermi edge on the eV scale instead of the $<$ 10 meV effects here. Furthermore, on a logarithmic scale, space charge effects can cause long high energy tails in the photoelectron spectrum \cite{Hellmann_PRB2009, Borgwardt_Thesis2016, Saathoff_PRA2008} that make it difficult to observe small signals from weakly excited samples. Here, with excellent harmonic isolation from our pulse-preserving monochromator and the absence of space charge effects, a precipitous drop of four orders of magnitude is observed in the EDC at the Fermi edge.   

\ind Next consider the red curve taken with the pump laser on. Before discussing the laser-induced features of the spectrum, consider first what is \emph{not} observed. The photoelectron spectrum is not shifted, broadened, or distorted due to space charge produced by the laser excitation as is commonly observed in pump probe experiments \cite{Ultstrup_JElecSpec2015, Oloff_JApplPhys2016, Borgwardt_Thesis2016, Saathoff_PRA2008}. This is because the high data rate enables the experiment to be performed under the low pump intensity of $1.3 \times 10^{9}$~W/cm$^2$. We measured the sample current from the pump excitation alone and found it to depend strongly on the region of sample probed, as in \cite{Saathoff_PRA2008}, but always at least one order of magnitude less than the XUV probe, or less than 22 electrons/pulse produced by the pump.

\ind The reflectivity of the gold sample is $>$ 97\% and the the laser-induced features in the EDC are dominated by the now well-known laser-assisted photoelectric effect (LAPE) \cite{Saathoff_PRA2008}. Briefly, LAPE is a dressing of the ionized electrons by the IR laser field producing sidebands at intervals of the photon energy \cite{Saathoff_PRA2008, Madsen_AmJPhys2005, Mahmood_NatPhys2016, Glover_PRL1996}. In Fig.~\ref{fig:TRUPS}a) a sideband of the surface state peak at 24.8 eV is observed 1.2 eV higher at 26 eV. Figure~\ref{fig:TRUPS}b) shows a measurement of the sideband amplitude at 26.0 eV kinetic energy as a function of the time delay between the IR pump and XUV probe. The data were accumulated in 10 minutes. Within statistical error, identical widths and time-zero positions are observed for cross correlations taken at both higher and lower kinetic energies, further confirming the LAPE mechanism, as hot electrons closer to the Fermi energy would show observable lifetimes \cite{FANN:1992uk,Cao:1998tp}.

\ind LAPE can be used to determine the time resolution of the instrument. A gaussian fit to the cross-correlation in Fig.~\ref{fig:TRUPS}b) gives a FWHM of 181 fs. The pump laser pulse duration at the sample position was not independently measured for this experiment, but at the output of the laser the pulse duration was measured to be $165 \pm 10$ fs and optimal compression gives 155 fs \cite{Li_RSI2016}. Taking the lowest possible value of the laser pulse duration then gives a conservative upper limit for the XUV pulse duration at the sample of $\sqrt{(181 \units{fs})^2 - (155 \units{fs})^2}$ = 93 fs. 

\ind Figure \ref{fig:TRUPS}c) shows amplitudes for the sideband at 26~eV kinetic energy for different pump pulse intensities obtained from fits to time-resolved scans as shown in Fig.~\ref{fig:TRUPS}b). The sideband amplitude is observed to be linear in the laser intensity with a slope of $1.34\pm0.13\times10^{-12}$ cm$^2$/W, and in excellent agreement with theory ($1.3\times10^{-12}$ cm$^2$/W) for our laser and experimental geometry, as calculated in appendix~\ref{ap:LAPE}. Even with the modest scan times of 10 minutes used to acquire this data, sideband amplitudes as small as $6\times10^{-4}$ are easily observed. We also note that with a multichannel electron analyzer, the delay-dependence for the full energy window of Fig.~\ref{fig:TRUPS}a) (or larger) could be obtained in parallel with no increase in data acquisition time.

\section{Conclusions}\label{sec:conclusions}

In this article, we have described ultrafast time-resolved photoemission experiments using a high-flux, high repetition rate, tunable XUV light source. The absence of space-charge effects in photoelectron spectra taken with nano-Ampere photocurrents have allowed us to observe LAPE with sideband amplitudes in the $10^{-3}$ to $10^{-4}$ range, orders of magnitude smaller than typical \cite{Saathoff_PRA2008, Tao_Science2016, Mathias_NatComm2016}. To our knowledge, these are the smallest LAPE signals observed, but more importantly they mimic the response of weakly excited samples. Being able to observe changes in the photoelectron spectra at this level enables studying samples excited weakly at low excitation fluences less than 10 $\mu$J/cm$^2$, which is typical for the more sensitive techniques of optical spectroscopy \cite{Huang_NatPhot2013}, two-photon photoemission \cite{Chan_Science2011_SingletFission}, and laser-based ARPES using 6 eV probe light \cite{Ishida_RSI2016}, but extremely difficult using space-charge limited HHG systems.

\ind Whereas space-charge sets a fundamental limit on most HHG-based photoemission instruments, the current time and energy resolution of the system do not represent any inherent limitations of the frequency-comb based methods used here, and are instead limited simply by the laser pulse duration and energy analyzer performance. Both of these are straightforward to improve. For example, in our setup sub-100 fs resolution could be obtained leaving the XUV probe arm unchanged and implementing nonlinear pulse compression in the pump arm \cite{Eidam_AppPhysB2008, Seidel_SciRep2017}, which has no demands on the temporal coherence of the pulse train. CE-HHG can also be performed with shorter driving pulses \cite{Pupeza_NatPhot2013, Lilienfein_OptLett2017}, if desired. Single-grating pulse preserving monochromators have been shown to be compatible with temporal resolutions as small as 8 fs \cite{Gierz_PRL2015}.

\ind Returning to Fig.~\ref{fig:sourcecomparison}, we use the conservative upper limit of 65 meV to compare the performance of our system against previous time-resolved photoemission work \cite{Mills_CLEO2017,*Jones_Private,Buss_SPIE2017,Frietsch_RSI2013,Artemis_Private,Chiang_NJP2015,Plotzing_RSI2016,Ojeda_SD2016,EICH_JElecSpec2014}. As can be seen, the present work enables a dramatic improvement over space-charge limited systems operating at lower repetition rate. Notable also is the work of Jones and co-workers \cite{Mills_CLEO2017,*Jones_Private} who have demonstrated to our knowledge the best resolution in HHG-based ARPES using a fixed photon-energy CE-HHG platform based on grating output coupling \cite{Yost_OptLett2008}. However, the grating output coupling method makes dynamic harmonic selection difficult, and also introduces larger pulse front tilts than the pulse-preserving monochromator, resulting in larger focused spot sizes and XUV pulse durations. 

\ind In the current experimental setup, the single-channel energy analyzer, which measures only one angle and energy at a time, currently sets the primary limit on the data rate. Even with this limitation, space-charge free time-resolved photoelectron spectra with high dynamic range can be recorded in minutes. Upgrading the electron analyzer to a multi-channel version will allow frames of time-resolved ARPES measurements to be accumulated at rates comparable to synchrotron beamlines. Notably, the \frep repetition rate is well suited to the recent advance of time-of-flight momentum microscopy developed by Sch\"onhense and coworkers \cite{Schonhense_JElecSpec2015}.

\section{Acknowledgements}

This work was supported by the U.S. Air Force Office of Scientific Research under Award No. FA9550-16-1-0164, the U.S. Dept. of Energy, Office of Science, Office of Basic Energy Sciences, under Award No. DE- SC0016017, and the Stony Brook Foundation Discovery Prize. MDK, ARM and MGW were supported by the U.S Department of Energy, Office of Science, and supported by its Division of Chemical Sciences, Geosciences, and Biosciences within the Office of Basic Energy Sciences under Contract No. DE-SC0012704. We thank Arthur K. Mills and David J. Jones at the University of British Columbia for many generous and fruitful discussions and T. C. Weinacht at Stony Brook for encouragement regarding LAPE. We are also grateful for useful data regarding space-charge effects in photoemission experiments from Robert A. Kaindl at Lawrence Berkeley National Laboratory and Cephise Cacho at the United Kingdom's Central Laser Facility.

\appendix
\section{Comparison of Sources in Figure 1}
\label{ap:fig1}

Both the development of HHG sources and their application to surface photoemission have been very active fields of research \cite{Mills_CLEO2017,*Jones_Private,Buss_SPIE2017, Wang_NatCom2015, Frietsch_RSI2013,Artemis_Private,Chiang_NJP2015,Plotzing_RSI2016,Ojeda_SD2016,EICH_JElecSpec2014,Hadrich_JPhysB2016}. To our knowledge, this work represents the first tunable HHG source applied to photoemission with nA sample currents capable of achieving sub-100 meV resolution in a spot size less than 1 mm. In Fig.~\ref{fig:sourcecomparison}, we have attempted to compare this work to previous published efforts applying HHG to sub-ps time-resolved surface photoemission. A time resolution of 1 ps corresponds to a 2 meV Fourier limit to the energy resolution. The following criteria were made in choosing which results to include: (1) Only systems with photon energy greater than 10 eV were included. (2) Only systems capable of achieving more than 1 pA of sample current were included. (3) Only systems driven by lasers with sub-ps pulse duration, such that they could in principle perform sub-ps time-resolved experiments, were included. (4) Only systems with published reports of being applied to photoemission have been included.

\ind For the energy resolution of photoemission experiments, several authors have shown that space charge broadening adds in quadrature with the photon energy bandwidth \cite{Plotzing_RSI2016, Hellmann_PRB2009}: 
\begin{equation}\label{eqn:scquad}
	\Delta E = \sqrt{ \Delta E_{\textss{b}}^2 + \Delta E_{h\nu}^2}
\end{equation}
with $\Delta E_{\textss{b}}$ the broadening due to space-charge and $\Delta E_{h\nu}$ the photon energy bandwidth. The black dashed lines in Fig.~\ref{fig:sourcecomparison} represent the asymptotes of Eq.~\ref{eqn:scquad} with $\Delta E_{h\nu}  \ll \Delta E_b$ . On the comparison plot, systems were categorized as not being space-charge limited (red edges) if the full flux can be applied to the sample in a 1 mm spot with $\Delta E_{\textss{b}}$ according to Eq.~\ref{eqn:sc} at least one order of magnitude less than $\Delta E_{h \nu}$. For non-space-charge limited experiments, the y-ordinate is determined either from published operating sample currents or published photon fluxes at the sample assuming a photoelectric yield of 0.1 electrons/photon. The x-ordinate is the published photon energy bandwidth. 

\ind For space-charge limited systems (black edges), the (x,y) positions represent the case where the sample current is such that $\Delta E_{\textss{b}} = \Delta E_{h\nu}$, and the energy resolution is reduced by a factor of $\sqrt{2}$. Since achieving resolution equal to the photon energy bandwidth requires substantially reduced sample current, this represents a practical compromise. The y-ordinate is calculated $I_{\textss{sample}} = e f_{\textss{rep}} D \Delta E_{h \nu} / m_b$, with D = 1 mm. The x-ordinate is then $\sqrt{2}$ times the published photon energy bandwidth. For ARPES, experiments are usually run with spot sizes smaller than 1 mm \cite{He_RSI2016}, such that even lower sample currents than plotted on Fig.~\ref{fig:sourcecomparison} are required to maintain energy resolution. Furthermore, we comment that for real-space imaging with photoemission electron microscopy, the space-charge constraints are even more severe \cite{Marsell_AnndePhysik2013, Chew_APL2012}. 

\section{THEORY FOR LAPE AMPLITUDE}
\label{ap:LAPE}

The laser-assisted photoelectric effect can be considered the result of dressing of the free-electron wavefunction \cite{Saathoff_PRA2008, Madsen_AmJPhys2005, Glover_PRL1996}. In the limit of electron kinetic energies much larger than the dressing laser photon energy and pondermotive energy much less than the photon energy, the amplitude of the $n$th sideband ($A_n$) becomes (in atomic units)

\begin{equation} \label{eqn:sidebandBessel}
	A_n = J_n^2 \left( \frac{\vecbf{p} \cdot \vecbf{E_0}}{\omega^2} \right) 
\end{equation}

where $\vecbf{p}$ is the vector momentum of the electron, $\vecbf{E_0}$ the laser electric field vector amplitude at the surface, and  $\omega$ is the laser frequency. The geometry of the experiment influences observed sideband amplitudes in two ways. First, energy can only be transferred between the laser field and the electron at the surface, such that for metallic surfaces only the component of the electric field normal to the surface contributes to LAPE \cite{Saathoff_PRA2008}. Second, due to the dot product in Eq.~\ref{eqn:sidebandBessel}, only the component of the electric field along the detection direction contributes. For our geometry, with the sample oriented 45 degrees to incident p-polarized beam and electrons detected along the surface normal, these factors are one and the same, and in the limit that the argument of the Bessel function in (\ref{eqn:sidebandBessel}) is much less than one we have (in SI units):

\begin{equation}\label{eqn:sidbandX}
	A_1 \approx \frac{4 \pi \alpha I E_{\textss{kin}}}{m_e \hbar \omega^4} \cos^2{45^{\circ}}
\end{equation}

where $\alpha$ is the fine structure constant, $I$ is the laser intensity ignoring the effects of the surface on the laser electric field, $E_{\textss{kin}}$ is the kinetic energy of the electron, $m_e$ is the mass of the electron, and $\hbar$ is Planck's constant. In the experiment, the laser intensity at which free electrons are generated varies in both space and time due to the finite extent of the XUV and laser beams. The observed sideband amplitude will thus be the space-time average
\begin{eqnarray}
	\left< A_1 \right> &=& \frac{2 \pi \alpha E_{\textss{kin}}}{m_e \hbar \omega^4} I_{\textss{peak}} \times \\ \nonumber
				&    & \int_{-\infty}^{\infty} dx \int_{-\infty}^{\infty} dy \int_{-\infty}^{\infty} dt G_{\textss{laser}}(x,y,t) G_{\textss{XUV}}(x,y,t) 
\end{eqnarray}

where $I_{\textss{peak}}$ is the peak intensity of the laser, $G_{\textss{laser}}(x,y,t)$ is a 3D Gaussian envelope function for the incident laser beam with unit amplitude and $G_{\textss{XUV}}(x,y,t)$ is a normalized 3D Gaussian envelope function for the XUV beam. Evaluating the space-time overlap integral with $1/e^2$ radii $w_{x,\textss{laser}} = 150$ $\mu$m, $w_{y,\textss{laser}} = 150$ $\mu$m, $w_{x,\textss{XUV}} = 49$ $\mu$m, $w_{y,\textss{XUV}} = 85$ $\mu$m, and FWHM pulse durations $T_{\textss{laser}} = 165$ fs and $T_{\textss{XUV}} = 93$ fs, the integral is 0.72 and the theoretical estimate for the observed sideband amplitude is $\left< A_1 \right>  = 1.3 \times 10^{-12} \times I_{\textss{peak}} \units{[W/cm}^2]$.


%

\end{document}